\documentclass[showpacs,amsmath,twocolumn,amssymb]{revtex4}
\usepackage{graphicx}
\usepackage{dcolumn}
\usepackage{bm}

\begin{document}
\title{A phenomenology analysis of the tachyon warm inflation in loop quantum cosmology}
\author{Kui Xiao}
\email{87xiaokui@mail.bnu.edu.cn}
\author{Jian-Yang Zhu}
\thanks{Author to whom correspondence should be addressed}
\email{zhujy@bnu.edu.cn}
\affiliation{Department of Physics, Beijing
Normal University, Beijing 100875, China}
\begin{abstract}
We investigate the warm inflation condition in loop quantum
cosmology. In our consideration, the system is described by a
tachyon field interacted with radiation. The exponential potential
function, $V(\phi)=V_0 e^{-\alpha\phi}\label{exp-p}$, with the same
order parameters $V_0$ and $\alpha$, is taken as an example of this
tachyon warm inflation model. We find that, for the strong
dissipative regime, the total number of e-folds is less than the one
in the classical scenario, and for the weak dissipative regime, the
beginning time of the warm inflation will be later than the tachyon
(cool) inflation.
\end{abstract}

 \maketitle

\section{Introduction}
The inflation is a very important concept in the modern cosmology
\cite{Liddle-book}. The standard model of the inflation was
introduced by Guth \cite{Guth-in}. However, because this model
relies on a scalar field which has no interaction with any other
fields, so that it is impossible that the radiation to be produced
during the inflation. This leads to a thermodynamically supercooled
phase of the Universe \cite{Berera-Warm}. So this standard
inflationary model needs a "graceful exit" to ensure the Universe
enters a radiation-dominated phase. In fact, it is not the only way
to describe the inflationary dynamics. Another model of the
inflationary picture is called the warm inflation
\cite{Berera-Warm}, as opposed to the conventional cool inflation.
In this model, the dissipative effects are very important during the
inflationary era, so the radiation is produced concurrently with an
inflationary expansion and there is no a separate reheating phase.
Also, the density fluctuation in the warm inflation arises from the
thermal fluctuation, rather than the vacuum fluctuation which
dominates the supercooled case \cite{Berera-Fang}. The radiation
dominates immediately as soon as the warm inflation ends. The matter
components of the Universe are created by the decay of either the
remaining inflationary field or the dominant radiation field
\cite{Berera-1996PRD}.

The warm inflation has been studied by many authors not only in
classical cosmology scenario but also in quantum cosmology scenario
(see  \cite{Berera-Warm} and references therein, and
\cite{Herrera-tach,Herrera-quan}). In this paper, we focus on a
tachyon warm inflation in loop quantum cosmology (LQC) scenario.

The application of loop quantum gravity techniques to homogeneous
cosmological models is known as LQC
\cite{Bojowald-1-11,Ashtekar-1-12,Singh-1-13}. Owing to the
homogeneity and isotropy of the spacetimes, the connection is
determined by a single parameter called $c$ and the triad is
determined by $p$. The variables $c$ and $p$ are canonically
conjugate with Poisson bracket $\{c,p\}=\gamma\kappa/3$, in which
$\gamma$ is the Barbero-Immirzi parameter and $\kappa=8\pi G$. In
the LQC scenario, the initial singularity is instead by a bounce.
Thanks to the quantum effect, the Universe is in an initially
contracting phase with minimal but not zero volume, and then the
quantum effect drives it to the expanding phase. And, in the
effective LQC scenario, the loop quantum effects can be very well
described by a effective modified Friedmann dynamics
\cite{Singh-PRD,Taveras-PRD}. There are two types of modifications,
one is the inverse volume correction, the other is the holonomy
correction. This paper we just discuss the holonomy correction. In
this effective LQC scenario, a factor of $(1-\rho/\rho_c)$ is added
to the standard Friedmann equation. For the correction term,
$-\rho/\rho_c$, comes with a negative sign, the Hubble parameter
$H$, and $\dot{a}$ vanishes when $\rho=\rho_c$, consequently the
quantum bounce occurs.

The warm inflation in the LQC scenario is considered by Herrera
\cite{Herrera-LQC} recently. The author discussed the inflationary
phenomenon described by a scalar field coupled to radiation. In this
paper, we would like to consider a tachyon warm inflation in the LQC
scenario. The tachyon field might be responsible for the inflation
at the early stage and could contribute to some new form of dark
matter at late times \cite{Sen-JHEP}. The behavior of the tachyon
field in LQC was studied by \cite{Sen-PRD}, in which the author
considered the inverse volume modification and found that there
exists a super accelerated phase in the semiclassical region. (For
arbitrary matter, the Universe will enter a super accelerated phase,
this issue was first considered by \cite{Singh-CQG}.) The tachyon
field in LQC based on $\rho^2$ modification was studied by
\cite{Xiong-tach}. The authors found that the inflation could be
extended to the region where the classical inflation stops. In this
paper, we consider the tachyon field is interacting with radiation
during the inflationary stage. Just as many authors have pointed out
(see \cite{Samart-dy},ect) the dynamical behaviors of interacting
field in LQC are very different from the ones in classical
cosmology. The purpose of this paper is comparing the difference
between the tachyon warm inflation in LQC and the one in classical
cosmology, and also, we will compare the difference between the
tachyon warm inflation and the tachyon (cool) inflation in LQC.

The paper is organized as follows. We present in Sec. \ref{sec2} the
basic concepts of the tachyon warm inflation in the LQC scenario,
and discuss in Sec. \ref{sec3} the exponential potential as an
example of this tachyon warm inflation. We end this paper with some
conclusions and discussions in Sec. \ref{sec4} .

\section{Tachyon warm inflation in LQC}\label{sec2}

In the LQC scenario, the Friedmann equation is modified as
\cite{Ashtekar-1-16,Ashtekar-1-17}
\begin{eqnarray}
H^2=\frac{\kappa}{3}\rho\left(1-\frac{\rho}{\rho_c}\right)\label{Fri-eq},
\end{eqnarray}
with the total energy density $\rho$ and the critical density
$\rho_c$, and $\kappa=8\pi G$. In our consideration,
$\rho=\rho_\phi+\rho_\gamma$, where $\rho_\phi$ denotes the energy
density of the tachyon field $\phi$, and $\rho_\gamma$ the radiation
energy density.

The dynamical equations for $\rho_\phi, \rho_\gamma$ in the warm inflation scenario are
\begin{eqnarray}
\dot{\rho}_\phi&=&-3H(\rho_\phi+p_\phi)-\Gamma\dot{\phi}^2,\label{drp}\\
\dot{\rho}_\gamma&=&-4H\rho_\gamma+\Gamma\dot{\phi}^2,\label{drg}
\end{eqnarray}
where the dot means the derivation with respect to time, and
$\Gamma$ is the dissipation coefficient responsible for the decay of
energy density of the tachyon field into radiation during the
inflationary era. $\Gamma$ can be considered as a constant or a
function of the field $\phi$, or the temperature $T$, or both of
them \cite{Herrera-tach}; and, according to the second law of
thermodynamics, $\Gamma>0$ should be hold. In this paper, for
simplicity we only consider to be a constant. The energy density
$\rho_\phi$ and the pressure $p_\phi$ of the tachyon field can be
written as \cite{Sen-PRD}
\begin{eqnarray}
\rho_\phi=\frac{V(\phi)}{\sqrt{1-\dot{\phi}^2}},\qquad
p_\phi=-V(\phi)\sqrt{1-\dot{\phi}^2}, \label{rp}
\end{eqnarray}
in which $V(\phi)$ is the potential of the tachyon field.

Considering Eqs.(\ref{drp}) and (\ref{rp}), one can get the equation
of motion (EoM) of the tachyon field
\begin{eqnarray}
\frac{\ddot{\phi}}{1-\dot{\phi}^2}+3H\dot{\phi}+\frac{V_{,\phi}}{V(\phi)}
=-\frac{\Gamma\dot{\phi}}{V(\phi)}\sqrt{1-\dot{\phi}^2},\label{EoM}
\end{eqnarray}
in which $V_{,\phi}=\frac{d V(\phi)}{d\phi}$.

In the LQC scenario, the condition for a bounce is $H=0$ and
$\dot{H}>0$, in which
\begin{eqnarray}
\dot{H}=-\frac{\kappa}{6}\left[3\rho(\phi)\dot{\phi}^2+4\rho_\gamma\right]
\left(1-2\frac{\rho_\phi+\rho_\gamma}{\rho_c}\right).\label{dH}
\end{eqnarray}
This means that the Universe will enter a super-inflation phase
immediately after bouncing. This is the first stage of the
inflation. The Hubble parameter $H$ will be increasing in this
stage. This stage is purely cased by the quantum effect and is very
short \cite{Chiou-inf}. We don't consider it in this paper.

According to Eqs.(\ref{Fri-eq}) and (\ref{dH}), the Raychaudhuri
equation can be written as
\begin{eqnarray*}
\frac{\ddot{a}}a &=&\dot{H}+H^2=H^2\left( 1-\varepsilon \right)  \\
&=&-\frac \kappa 6\left\{ \rho \left( 1-\frac \rho {\rho _c}\right)
+3\left[ p\left( 1-\frac{2\rho }{\rho _c}\right) -\frac{\rho
^2}{\rho _c}\right] \right\} ,
\end{eqnarray*}
with the total pressure $p=p_\phi+p_\gamma$ and the slow-roll
parameter $\varepsilon=-\frac{\dot{H}}{H^2}$. Inflation is often
defined as a period of accelerated expansion, i.e., $\ddot{a}>0$. In
this paper, we will focus on the evolution of the field in the
slow-roll inflationary era. During this era, the potential dominates
over the kinetic energy of the tachyon field, i.e. $\rho_\phi\simeq
V(\phi)$,  and the energy density of the radiation, i.e.,
$V(\phi)>\rho_\gamma$. We can also assume that $\dot{\phi}^2\ll 1$
and $\ddot{\phi}\ll [3H+\Gamma/V(\phi)]\dot{\phi}$ in this region.
Then, the Friedmann equation is reduced to
\begin{eqnarray}
H^2=\frac{\kappa}{3}V(\phi)\left(1-\frac{V(\phi)}{\rho_c}\right).
\label{rFri}
\end{eqnarray}
And the EoM of the tachyon field (\ref{EoM}) becomes
\begin{eqnarray}
3H(1+R)\dot{\phi}=-\frac{V_{,\phi}}{V(\phi)}, \label{rEoM}
\end{eqnarray}
in which $R$ is the rate defined as
\begin{eqnarray}
R=\frac{\Gamma}{3HV(\phi)}. \label{rate}
\end{eqnarray}
For a strong (weak) dissipative regime, we have $R\gg 1$ ($R<1$),
i.e., $\Gamma\gg 3HV$ ($\Gamma<3HV$).

The Raychaudhuri equation can be rewritten as
\begin{eqnarray}
\frac{\ddot{a}}{a}=-\frac{\kappa}{6}\left(\rho_{\rm{eff}}+3p_{\rm{eff}}\right)\label{mod-Ray},
\end{eqnarray}
with
\begin{eqnarray}
\rho_{\rm{eff}}=\rho_\phi\left(1-\frac{\rho_\phi}{\rho_c}\right),\quad
p_{\rm{eff}}
=p_\phi\left(1-\frac{2\rho_\phi}{\rho_c}\right)-\frac{\rho_\phi^2}{\rho_c}.\label{eff-rho
p}
\end{eqnarray}
The inflation ends when $\ddot{a}=0$, this implies that $\rho_{\rm{eff}}+3p_{\rm{eff}}=0$.
So, at the point of ending of the inflation, one has
\begin{eqnarray}
\rho_\phi=\frac{3\omega+1}{3\omega+4}\rho_c,
\end{eqnarray}
with the equation of state parameter of the tachyon field
$\omega=p_\phi/\rho_\phi$. In the classical cosmology,
$\rho\ll\rho_c$, it is easy to find that the inflation ends when
$\rho_\phi=-3p_\phi$, i.e., $\omega=-\frac13$ (see
Eq.(\ref{mod-Ray}) and consider $\frac{\rho_\phi}{\rho_c}=0$). But
in the LQC scenario, if $\omega=-\frac13$, $\rho_\phi=0$, this means
the energy density of the tachyon field should be zero at the
inflation ending point. But it is easy to verify that $\rho_\phi>0$
when $\omega=-\frac13$ (see Eq.(\ref{rp})). This means that the
inflation phase still exists in LQC while the classical inflation
stops. Notice that we suppose the quantum effect can not be ignored.
If $\rho_\phi\ll \rho_c$ where the quantum effect can ge ignored,
just as \cite{Xiong-tach} argued, the quantum and the classical
inflation have the same trajectory. This phenomena is as same as the
tachyon (cool) inflation in LQC. This is not surprise. For the
energy density of the tachyon field (or the potential of the tachyon
field) dominates over the energy density of radiation.

Also, as the condition is in the classical tachyon warm inflation
\cite{Herrera-tach}, we can consider that the radiation production
is qusi-stable during the warm inflation region. Then the energy
density of radiation can be reduced as
\begin{eqnarray}
\rho_\gamma=\frac{\Gamma\dot{\phi}^2}{4H}.\label{rg}
\end{eqnarray}
Considering Eqs.(\ref{rFri}), (\ref{rEoM}) and (\ref{rg}), one can
obtain
\begin{eqnarray}
\rho_\gamma=\frac{R}{4\kappa(1+R)^2}\frac{V_{,\phi}^2}{V(\phi)^2(1-V(\phi)/\rho_c)}.\label{rg-1}
\end{eqnarray}

Under those conditions, one can get the slow-roll parameter
\begin{eqnarray}
\varepsilon=\frac{1}{2\kappa}\frac{1}{V(\phi)
(1+R)}\left(\frac{V_{,\phi}}
{V(\phi)}\right)^2\frac{[1-2V(\phi)/\rho_c]}{[1-V(\phi)/\rho_c]^2}.\label{vare}
\end{eqnarray}
The last fraction is caused by the modification of the quantum geometry.  Also,
we can rewrite $\varepsilon$ as a function of $\rho_\gamma,\rho_\phi$ and the rate $R$:
\begin{eqnarray}
\varepsilon=\frac{\rho_\gamma}{\rho_\phi}\frac{2(1+R)}{R}\frac{1-2\rho_\phi/\rho_c}{1-\rho_\phi/\rho_c}.
\end{eqnarray}
The accelerated expansion occurs if $\varepsilon<1$, i.e.,
$\ddot{a}>0$. Then, the relationship between $\rho_\phi$ and
$\rho_\gamma$ in the accelerated region is
\begin{eqnarray}
\frac{2(1+R)}{R}\rho_\gamma<\frac{1-\rho_\phi/\rho_c}{1-2\rho_\phi/\rho_c}\rho_\phi.
\end{eqnarray}
The inflation ends when the slow-roll conditions are violated, i.e.,
$\varepsilon=1$, which implies
\begin{eqnarray}
\left(\frac{V_{,\phi}}{V(\phi)}\right)^2\frac{[1-2V(\phi)/\rho_c]}{\kappa
V(\phi)[1-V(\phi)/\rho_c]^2}= 2(1+R).
\end{eqnarray}
The number of e-folds before inflation ends is
\begin{eqnarray}
N(\phi)=\kappa\int^{\phi_i}_{\phi_f}\frac{V(\phi)^2[1-V(\phi)/\rho_c]}{V_{,\phi}}(1+R)d\phi.
\label{e-folds}
\end{eqnarray}
in which $\phi_i,\phi_f$ denote the values of the tachyon field at
the beginning and the end of inflation respectively.

Comparing above equations with the ones in the classical scenario
\cite{Herrera-tach}, one can find that the term modified by quantum
geometry is very important in the inflationary regions. Notice that
we assume the inflation happens very near the quantum dominated
region. If the inflation happens far away from the quantum dominated
region, then the quantal modified term $\frac{\rho}{\rho_c}=0$ and
the variables which are shown by above equation are as same as the
ones in the classical cosmology. As an example, we will discuss the
warm inflation of the tachyon field with an exponential potential in
strong and weak dissipative regime.

\section{An example: exponential potential}\label{sec3}

At the Sec. \ref{sec2} we get the expressions for the variables of
the tachyon warm inflation with general potential. As an example, we
consider in this section the exponential potential \cite{Sen-MPLA}
\begin{eqnarray}
V(\phi)=V_0 e^{-\alpha\phi}\label{exp-p},
\end{eqnarray}
with the constant $V_0>0$ and the tachyon mass $\alpha>0$ (with unit
$m_p$). The discussion will be concerned with the strong and weak
dissipative regime. For simplicity, we just consider
$\Gamma=\Gamma_0$ is a constant.

\subsection{Warm inflation in the strong dissipative regime}\label{sec3.1}
For a strong dissipative regime, we have $R\gg 1$, i.e., $\Gamma\gg
3HV$. Considering Eqs.(\ref{rEoM}) and (\ref{exp-p}), one can get
\begin{eqnarray}
\Gamma \dot{\phi}=V_0\alpha e^{-\alpha\phi}.
\end{eqnarray}
Here, we have taken into account the condition of $\Gamma\gg 3HV$.
And from now on, we will just consider $\Gamma$ as a constant
dissipation coefficient, i.e., $\Gamma=\Gamma_0$. Then, one can
obtain the evolution of the tachyonic field:
\begin{eqnarray}
\phi(t)=\frac{1}{\alpha}\ln\left(\frac{\alpha^2
V_0}{\Gamma_0}t+e^{\alpha\phi_i}\right),\label{phi-t}
\end{eqnarray}
in which $\phi_i$ is the initial value of $\phi$ at the time that
the slow-roll inflation begins. It is straightforward to see that
$\phi(t)$ is not the function of the quantum geometry correction
$\left(1-V(\phi)/\rho_c\right)$. This is because we consider
$\Gamma\gg 3HV$, then the Hubble parameter is ignored when we
consider the EoM of the field. This is as same as the evolution in
the tachyon warm inflation of the classical cosmology scenario
\cite{Herrera-tach}, but is different from the standard inflation of
tachyon field in LQC \cite{Xiong-tach} in which the correction came
from the quantum geometry effects plays an important role.

To get an explicit expression of the number of e-folds, we will
resort to the values of the potential at the beginning and the end
of the inflation. According to Eq.(\ref{e-folds}), we can get
\begin{eqnarray}
N_{Strong}(V)=\frac{\Gamma_0}{\alpha^2}\sqrt{\frac{3}{\kappa}}\int^{V_i}_{V_f}\frac{\sqrt{1-V/\rho_c}}{V^{3/2}}dV,
\label{N-V}
\end{eqnarray}
in which $V_i, V_f$ denote the values of the potential of the
tachyon field at the beginning and the end of the the inflation
respectively. The inflation ends when $\ddot{a}=0$, i.e.,
$H^2_f=-\dot{H}_f$. This implies that
\begin{eqnarray}
\frac{\kappa}{3}V_f\left(1-\frac{V_f}{\rho_c}\right)=\frac{\kappa}{2}V_f\dot{\phi}^2_f\left(1-\frac{2V_f}{\rho_c}\right)=\frac{\kappa^2}{2\Gamma^2_0}V^3_f
\left(1-\frac{2V_f}{\rho_c}\right).\nonumber
\end{eqnarray}

We have used Eqs.(\ref{dH}), (\ref{rFri}), (\ref{rEoM}) and
(\ref{rate}), and the slow-roll approximate $\rho_\phi\simeq
V(\phi), \rho_\phi\gg\rho_\gamma$. Solving the above equation, one
can get
\begin{widetext}
\begin{eqnarray}
V_f=V(\phi=\phi_f)&=&\frac16{\frac {\sqrt
[3]{-30\rho_{{c}}\alpha{\Gamma_0 }^{2}+{\rho_{{c}
}}^{3}{\alpha}^{3}+2\Gamma_0\sqrt {-16{\Gamma_0 }^ {4}+213{\Gamma_0
}^{2}{\rho_{{c}}}^{2}{\alpha}^{2}-18\,{\rho_{{c}}}^{4
}{\alpha}^{4}}}}{\alpha}}\nonumber\\
&&+\frac16{\frac {4{\Gamma_0 }^{2}+{\rho_{{c}}}^{
2}{\alpha}^{2}}{\alpha\sqrt [3]{-30\,\rho_{{c}}\alpha\,{\Gamma_0 }^{2}
+{\rho_{{c}}}^{3}{\alpha}^{3}+2 \Gamma_0\sqrt {-16
{\Gamma_0 }^{4}+213\,{\Gamma_0 }^{2}{\rho_{{c}}}^{2}{\alpha}^{2}-18{
\rho_{{c}}}^{4}{\alpha}^{4}}}}}+\frac16\rho_{{c}}.\label{V-f}
\end{eqnarray}
Obviously, this result depends on the values of $\Gamma_0,\alpha$ and $\rho_c$.

Integrating Eq.(\ref{N-V}), one can get
\begin{eqnarray}
N_{Strong}(V)
=\left.\frac{\Gamma_0}{\alpha^2}\sqrt{\frac{3}{\kappa}}\left[2\,{\frac
{\rho_{{c}}-V}{\sqrt {V \left(\rho_{{c}}-V \right) \rho_
{{c}}}}}+\arctan \left( {\frac {\sqrt {\rho_{{c}}} \left(
-1/2\,\rho_{ {c}}+V \right) }{\sqrt
{-\rho_{{c}}{V}^{2}+{\rho_{{c}}}^{2}V}}}\right) {\frac {1}{\sqrt
{\rho_{{c}}}}} \right]\right|_{V_i}^{V_f}.
\end{eqnarray}
\end{widetext}

The number of e-folds depends on $V_i=V_0\exp^{-\alpha\phi_i}$ and
$V_f$ given in Eq.(\ref{V-f}). To ensure $V_f$ is a real number,
$-16 {\Gamma_0 }^{4}+213{\Gamma_0
}^{2}{\rho_{{c}}}^{2}{\alpha}^{2}-18{ \rho_{{c}}}^{4}{\alpha}^{4}>0$
should be held. This gives a constraint on $\Gamma_0,\alpha$, i.e.,
$0.32<\frac{\Gamma_0}{\alpha}<2.25$. This means the same order of
the magnitude of $\Gamma_0, \alpha$. Note that this order of the
magnitude of $\alpha$ is smaller than the one who used in the
tachyon (cool) inflation \cite{Xiong-tach}. Always, the
observational data gives a constraint on the value of $\Gamma_0$,
just as the jobs of \cite{Herrera-tach}. This constraint connects
with the spectrum of the scalar perturbations and the tensor-scalar
ratio. But unfortunately, the holonomy correction to the scalar
perturbation is still incomplete, even for the scalar field. So, the
value of $\Gamma_0,\alpha$ is still need for more research.

If the quantum correction can be ignored, the total e-folding number
will less than 1 when one discusses the strong dissipation in the
classical cosmology \cite{Herrera-tach}. To compare the total
e-folding number in LQC and the one in classical cosmology, we can
employ the slow roll condition and approximately replace the term
$\left( 1-V\left( \phi \right) /\rho _c\right) $ by a constant
$\lambda <1$ \cite{Xiong-tach}. Under this approximation, the Hubble
parameter can be rewritten as
\begin{eqnarray}
\frac{\dot{a}}{a}\simeq \sqrt{\frac{\kappa}{3}\lambda V(\phi)}.
\end{eqnarray}

Considering Eqs.(\ref{exp-p}) and (\ref{phi-t}), one can get
\begin{eqnarray}
\frac{a}{a_i}=\exp\left[\frac{2\Gamma_0}{V_0\alpha}\sqrt{\frac{\kappa\lambda
V_0}{3}}\left(\sqrt{\frac{\alpha^2
V_0}{\Gamma_0}t+e^{\alpha\phi_i}}-
e^{\frac12\alpha\phi_i}\right)\right]\label{s-f},
\end{eqnarray}
with the initial value $a_i$ of $a$ at the time at the beginning of
the inflation $t_i$. For $0<\lambda<1$, the scale factor in this
region is smaller than the one in classical scenario
\cite{Herrera-tach}. Based on Eq.(\ref{s-f}), one can get the time
at the end of the inflation $t_f$, i.e., the time of $\ddot{a}=0$.
\begin{eqnarray}
t_f=\frac{3\alpha^2}{4\kappa\lambda\Gamma_0}-\frac{\Gamma_0}{\alpha^2V_0}e^{\alpha\phi_i}.
\end{eqnarray}
And, this time is bigger than the one in classical scenario
\cite{Herrera-tach}. Inserting this equation into Eq.(\ref{s-f}),
one can obtain \cite{Sami-info}
\begin{eqnarray}
\frac{a_f}{a_i}=\exp\left[\frac{2\Gamma_0}{V_0\alpha}\left(\sqrt{\frac{\alpha^4V^2_0}{4\Gamma_0^2}-\frac{\kappa\lambda
V_0}{3}e^{ \alpha\phi_i}}-e^{\frac12 \alpha\phi_i}\right)
\right].\label{N-a}
\end{eqnarray}
It is easy to find that, for an exponential potential in slow-roll
limit, the number of e-folds of the tachyon warm inflation in LQC is
smaller than the classical one. This is caused by the quantum
correction. Then it is not surprise that this result is as same as
the ones in tachyon (cool) inflation \cite{Xiong-tach}. So, if we
believe the quantum effect cannot be ignored, then the e-folding
number will small than the one in classical cosmology. This means
that $N_{Strong}<1$ for the strong dissipative regime. The tachyon
field will become bigger and bigger during the slow-roll inflation,
then $V=V_0 e^{-\alpha\phi}$ will become smaller and smaller, so
$\lambda=(1-V/\rho_c)$ will become smaller and smaller during the
slow-roll inflation scenario. This means that $\frac{a_f}{a_i}$ will
be bigger than the one which is described by Eq.(\ref{N-a}). But the
total number of e-folds in LQC is smaller than the one in classical
cosmology. Notice that the total number of e-folds in LQC did not include the e-folds of super-inflation.

\subsection{Warm inflation in the weak dissipative regime}\label{sec3.2}
\begin{figure}[h!]
\begin{center}
\includegraphics[width=0.45\textwidth]{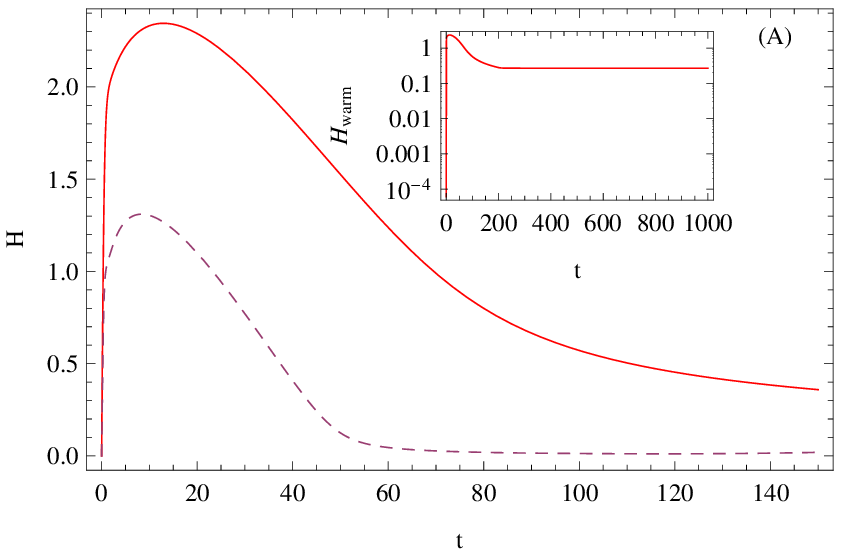}
\includegraphics[width=0.45\textwidth]{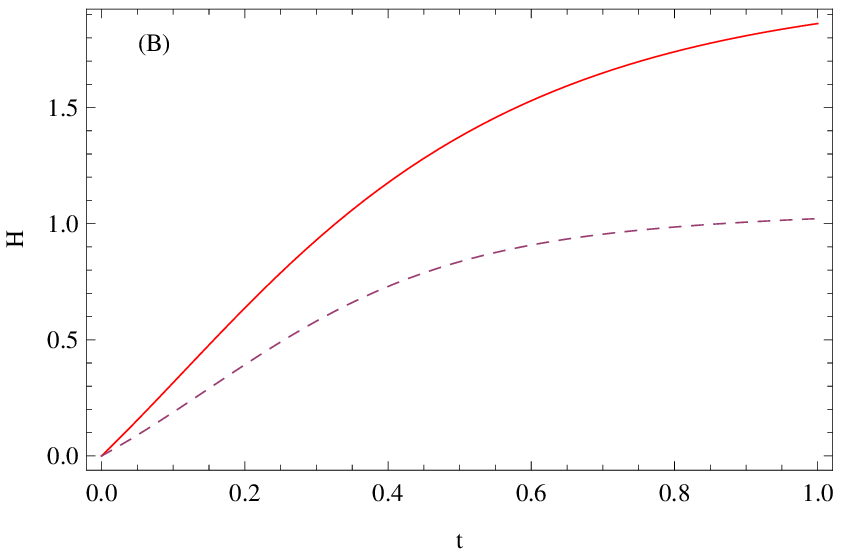}
\includegraphics[width=0.45\textwidth]{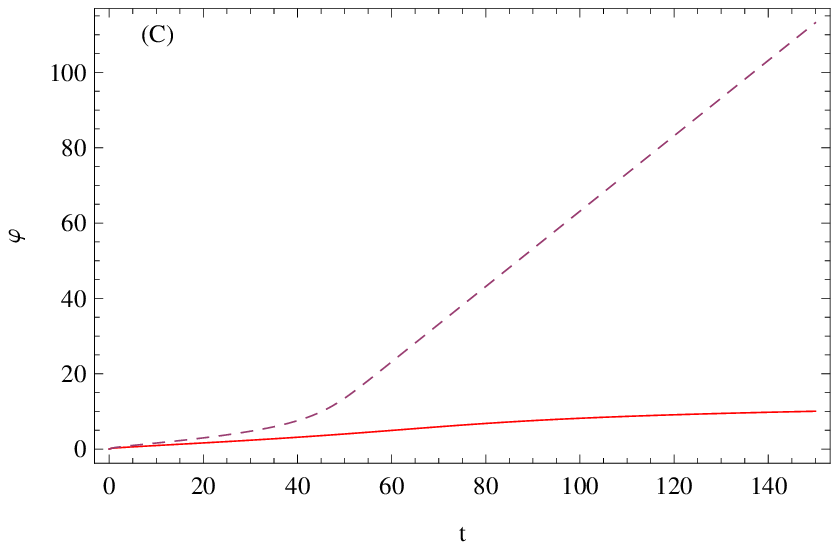}
\caption{(Color online). Evolutions for the case of the tachyon warm
inflation (solid curve) and the tachyon (cool) inflation (dashed
curve) with $V_0=0.82,\alpha=0.5,\Gamma=0.1$ and
$\rho_\gamma(t=0)=0$. (A). Evolution of the Hubble rate $H$. The
first period after bouncing is super-inflation ($H$ is increasing.
This period is shown in the (B)), and then $H$ decrease till the
inflation stage. The sub-picture in (A) is the evolution of $H$ in
warm inflation. (C). Evolution of the tachyon field $\phi$ with
exponential potential $V=V_0 e^{-\alpha\phi}.$ Unlike the scalar
field with the quadratic potential that it will enter an oscillatory
epoch, the tachyon field will always increase.}\label{Fig1}
\end{center}
\end{figure}

The tachyon warm inflation in the weak dissipative regime in
classical cosmology was discussed by \cite{Herrera-weak}.
Considering $\Gamma=const.$ and the weak dissipation, i.e., $R<1$,
the total number of e-folds can be rewritten as
\begin{eqnarray}
N_{Weak}(\phi)&=&\kappa\int^{\phi_i}_{\phi_f}\frac{V(\phi)^2[1-V(\phi)/\rho_c]}{V_{,\phi}}d\phi
\end{eqnarray}
It is obvious that $N_{Weak}(\phi)$ depends on the tachyon mass
$\alpha$ and the relationship between $V$ (this means that it also
dependents on $V_0$) and $\rho_c$. In \cite{Herrera-weak}, the
authors shown $V_0\sim 10^{-10} m_p^4$ and $\alpha\sim 10^{-6}m_p$.
These data are based on the WMAP five-year data and the Sloan
Digital Sky Survey (SDSS) large-scale structure surveys and the
perturbations of the tachyon field in the classical cosmology
scenario. If these data is still tenable in the LQC scenario, then
$V(\phi)\ll \rho_c$ for $\phi\gg -2.3\times 10^7$
($V(\phi)\simeq\rho_c$ for $\phi= -2.3\times 10^7$). Then the
quantum effect can be ignored. The total number of e-folds of the
tachyon warm inflation in the weak dissipative regime in LQC
scenario is as same as the one in the classical scenario. But as
same as we mentioned before, the perturbation theory of the tachyon
field in the LQC scenario is still need for more study, then we
cannot obtain $V_0, \alpha$ through the observe data. The main aim
of this subsection is comparing the difference between the tachyon
warm inflation and the tachyon (cool) inflation in LQC, we assume
$V_0=0.82,\alpha=0.5$. And these differences are shown in the
Fig.\ref{Fig1}.

Figure \ref{Fig1} shows that there are two stages of the inflation.
The first is a stage of the super-inflation near the bouncing epoch
as we discussed in the Sec.\ref{sec2}. It is easy to find that the
super-inflation ends very quickly. The second stage of the inflation
begins at the stage $H\simeq Const.$. This stage is far away from
the bouncing epoch and the quantum correction is completely
negligible. This is nothing but just the standard slow-roll
inflation.

The evolution pictures of two inflationary scenario have the same
directions when we consider the same initial condition
($\phi(t=0)=0,\dot{\phi}(t=0)=0,a(t=0)=1,\dot{a}(t=0)=0$). The
variable $H$ in warm inflation is bigger than the one in (cool)
inflation at the same time. This is reasonable. Since the total
energy density $\rho$ in warm inflation includes the energy density
of the tachyon field $\rho_\phi$ and the radiation $\rho_\gamma$,
but the one in (cool) inflation just includes $\rho_\phi$. We assume
these two different inflation models have the same initial
conditions. So $\rho>\rho_\phi$ at the same time. And we can find
that the field $\phi$ in warm inflation is smaller than the one in
(cool) inflation. This is because the energy density of the tachyon
field decays into the radiation during the inflationary era. Also,
we can see the non-inflationary phase (between the super-inflation
and the slow-roll inflation) in warm inflation is longer than the
one in (cool) inflation. This phase is an indirect loop quantum
gravity effect \cite{Chiou-inf}. It is easy to find that the
slow-roll inflation is beginning at $t\simeq 48$ in the (cool)
inflation but at $t\simeq 200$ in the warm inflation.

In this section, we discuss the warm inflation of the tachyon field
with an exponential potential in the LQC scenario. We find that the
total number of e-fold is less than 1 if we consider the strong
dissipative regime ($R \gg 1$). And if we consider the weak
dissipative regime ($R<1$), the beginning time of the slow-roll
inflation in warm inflation is later than the one in (cool)
inflation when we consider the same initial condition. But due to
the perturbation theory of loop quantum cosmology is still open, we
cannot get the parameters $V_0, \alpha$ through the observational
data. Therefore it is still impossible to get the special total
number of e-folds number.

\section{conclusions and discussions}\label{sec4}
As showing in Eq.(\ref{Fri-eq}), the Friedmann equation in LQC adds
a factor of $(1-\rho/\rho_c)$ in the right side of the standard
Friedmann equation. The correction term $\rho/\rho_c$ comes with a
negative sign, this makes it possible that $\dot{a}=0$ when
$\rho=\rho_c$, and the bounce occurs. At the bounce point, $H=0$ and
$\dot{H}$ is positive. The Universe enters a super-inflation stage.
(If one considers the inverse volume modification, the Universe will also enter
a super-inflation stage \cite{Singh-CQG}.)
Eq.(\ref{dH}) shows that $\dot{H}$ continues to positive till
$\rho=\frac12\rho_c$ (at which point $\dot{H}$ is vanishes, and
after this point, it will become negative.). Thus, every LQC
solution has a super-inflation phase from $\rho=\rho_c$ to
$\rho=\frac12\rho_c$. However, we must recognize that this stage is
very short, so that the super-inflation can not substitute for the
slow-roll inflation.

In this paper, we studied the tachyon warm inflation model in the
LQC scenario. At first, we considered the tachyon field with a
general potential coupled with radiation field in the slow-roll
inflation phase. During this inflationary era, the potential
dominates over the kinetic energy of the tachyon field and the
energy density of radiation. Then the modified Friedmann equation
and Rachaudhuri equation have the same expressions  with the tachyon
(cool) inflation in LQC. We found that the warm inflation phase in
LQC will expand to the region where the classical inflation stops.
The interacting term will modify the EoM of the tachyon field in the
slow-roll approximate. Then the energy density of radiation has also
been modified. Based on those conditions, we got a general
relationship between the tachyon field and radiation energy density,
and obtained the relationship between $\rho_\gamma,\rho_\phi$ in the
accelerated region. We found that the number of e-folds before
inflation ends depends on the modification term $(1-V(\phi)/\rho_c)$
and the rate $R$.

And then, as an example, we discussed the tachyon warm inflation
with an exponential potential in a strong and a weak dissipative
regime. For the strong dissipative regime ($\Gamma\gg 3HV$), the
quantum geometry effects did not change the evolution of the tachyon
field $\phi(t)$. But it will modify the energy density of radiation,
the scale factor, and the total number of e-folds. We found that the
total number of e-folds in LQC is less than the one in the classical
scenario if we just consider the slow-roll inflation phase. We also
discussed the difference between the tachyon warm inflation in the
weak dissipative regime and the tachyon (cool) inflation in LQC. We
found that the Hubble parameter $H$ in warm inflation is bigger than
the one in (cool) inflation at the same time, the beginning time of
slow-roll inflation in warm inflation is later then the one in
(cool) inflation, and the value of the tachyon field $\phi$ at the
beginning time of warm inflation is less than the one in (cool)
inflation.

\acknowledgements This work was supported by the National Natural
Science Foundation of China under Grant No. 10875012 and the
Fundamental Research Funds for the Central Universities.

\end{document}